# Commercial building decarbonization across countries: Future emission trajectories and mitigation priorities

Qinwen Tang [1], Tianyi Wang [2], Nan Zhou [3], Minda Ma [1,4*]

1. School of Architecture and Urban Planning, Chongqing University, Chongqing, 400045, P. R. China
2. School of Management Science and Real Estate, Chongqing University, Chongqing, 400045, P. R. China
3. Energy and Resources Group, University of California, Berkeley, CA 94720, United States
4. Lawrence Berkeley National Laboratory, Berkeley, CA 94720, United States

- Corresponding author: Dr. Minda Ma, Email: maminda@lbl.gov; minda.ma@cqu.edu.cn
Homepage: https://globe2060.org/MindaMa/

## Graphical abstract

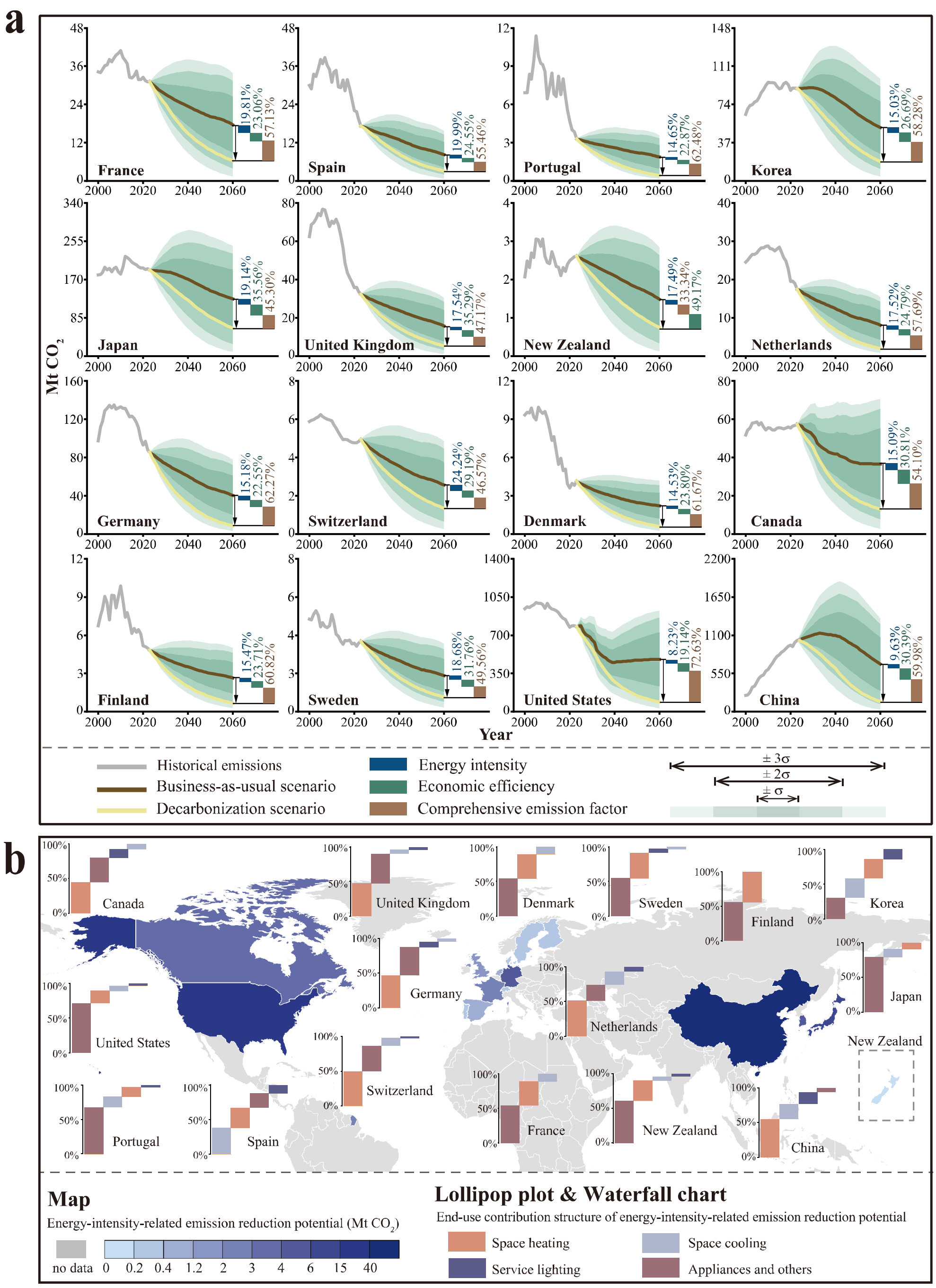

**Graphical abstract.** (a) Contributions of energy intensity, economic efficiency, and the comprehensive emission factor to the emission reduction potential of commercial building operations in 2060; (b) contributions of four end uses to energy-intensity-related emission reduction potential in commercial buildings in 2060. Note: End-use categories vary across countries because of data availability. Data on service lighting were unavailable for France, Denmark, and Japan. Only space heating and appliances and other end uses were included for Finland.

## Highlights

- Kaya-based modeling and Monte Carlo simulations project 16 countries' emissions to 2060.
- China's commercial building emissions could peak around 2033 under business as usual.
- Mitigation potential reaches 1123 $MtCO_2$ by 2060, led by China, the US, and Japan.
- Lower comprehensive emission factors dominate mitigation in most countries.
- Heating dominates China's efficiency potential; appliances lead in the US and Japan.

## Abstract

With energy-related carbon dioxide emissions remaining at high levels, decarbonizing commercial building operations across key economies is critical for achieving global climate targets. This study assesses future operational carbon trajectories, mitigation potential, and the driving factors of this potential in commercial building operations across 16 countries from 2024 to 2060 by combining a Kaya-based emission model and Monte Carlo uncertainty analysis. The key findings indicate that (1) under the business-as-usual scenario, China's commercial building operational carbon emissions are projected to peak at 1150 (±120) million tonnes of carbon dioxide ($MtCO_2$) in 2033 (±5 years), while the remaining countries exhibit differentiated carbon lock-in levels, with the highest level in the United States at 470 (±89) $MtCO_2$. (2) The total emission reduction potential increases from 352 $MtCO_2$ in 2030 to 1119 $MtCO_2$ in 2050 and stabilizes at 1123 $MtCO_2$ in 2060, with China, the United States, and Japan contributing 84.7%–88.9% of the total potential. (3) By 2060, the mitigation potential from energy intensity reduction is concentrated mainly in China and the United States, reaching 52 and 32 $MtCO_2$, respectively; end-use decomposition further identifies space heating and appliances as key targets for decarbonizing commercial building operations. In summary, this study characterizes future operational carbon trajectories and clarifies the drivers and end-use priorities of mitigation potential in commercial building operations, providing cross-country evidence for differentiated strategies for long-term commercial building decarbonization.

## Abbreviations notation

| | |
|---|---|
| BEMS | Building energy management system |
| $CO_2$ | Carbon dioxide |
| EKC | Environmental Kuznets curve |
| GDP | Gross domestic product |
| HVAC | Heating, ventilation, and air conditioning |
| IPAT | Impact of population, affluence, and technology |
| $MtCO_2$ | Million tonnes of carbon dioxide |
| SD | Standard deviation |
| °C | Degrees Celsius |

**Nomenclature**

| | |
|---|---|
| $C$ | Total operational carbon emissions from commercial buildings |
| $C^{dynamic}$ | Emission result after considering parameter uncertainty |
| $C^{static}$ | Emission result calculated from static scenario parameters |
| $E$ | Total energy consumption of commercial buildings |
| $E_i$ | Energy consumption of end use $i$ |
| $e_i$ | Energy intensity of end use $_i$ |
| $F$ | Commercial building floor area |
| $Gs$ | Service value added |
| $i$ | Economic efficiency |
| $K$ | Comprehensive emission factor |
| $P$ | Population of each country |
| $r$ | Per capita service value added |
| $T$ | Target year during 2024–2060 |
| $X$ | General parameter, representing P, r, i, ei, or K |
| $X^{dynamic}$ | Dynamic value of parameter X after considering parameter uncertainty |
| $X^{static}$ | Static value of parameter X |
| $\omega$ | Random parameter or random disturbance term |
| $\omega_X$ | Random disturbance term of parameter X |
| $\sigma$ | Standard deviation |

## 1. Introduction

Climate change remains a major challenge to global sustainable development [1], while energy-related carbon dioxide ($CO_2$) emissions remain high, reaching a record 37.8 billion tonnes in 2024 [2]. Against this background, the Paris Agreement calls for limiting the global temperature rise to well below 2 degrees Celsius (°C) above preindustrial levels and pursuing efforts to restrict the increase to 1.5°C [3]. The building sector contributes nearly 40% of total carbon emissions [4] and consumes more than 34% of global energy use [5], indicating that improving energy efficiency is an important part of emission reduction efforts. Commercial buildings represent a major component of the building sector [6]. In particular, commercial building operations account for approximately 10% of global carbon emissions and have considerable potential for deep decarbonization [7]. From a life-cycle perspective, operational emissions are the dominant contributor to building carbon emissions, accounting for approximately 78% of total life-cycle carbon emissions [8]. Therefore, reducing operational carbon emissions from commercial buildings is a key pathway for mitigating building-sector emissions.

Understanding the mitigation potential of commercial building operations requires moving beyond aggregate emission reductions to the factors that shape this potential across different decarbonization pathways. Existing research has explored emission reduction in buildings from different perspectives, including energy efficiency improvement, low-carbon energy supply, end-use system optimization, and building energy standards [9]. Some studies have also assessed building decarbonization pathways for specific countries, regions, or historical emission changes [10]. **However, cross-country evidence linking future operational emission trajectories, mitigation drivers, and end-use priorities in commercial buildings remains limited.** Therefore, this study proposes three research questions to examine future operational carbon trajectories, assess mitigation potential, and identify differentiated mitigation priorities across countries:

- What are the future operational carbon trajectories of commercial buildings?
- What are the emission reduction potentials and their driving factors?
- How do mitigation priorities vary across countries and end uses?

To answer these questions, this study develops a Kaya-based model for commercial building operations from 2024 to 2060, incorporating the effects of population, service-sector development, commercial floor area, energy intensity, and energy supply structure to analyze future operational carbon trajectories. To account for uncertainties in future emissions, a Monte Carlo simulation is further incorporated into the scenario framework to assess uncertainties in carbon peaks, peak time, and carbon lock-in levels across countries. On this basis, scenario simulation is used to quantify the future emission reduction potential of commercial building operations and to decompose its driving factors. Finally, this study discusses corresponding deep decarbonization strategies to inform policy decisions by governments and relevant institutions on reducing operational carbon emissions from commercial buildings.

**The main contribution of this study lies in an integrated framework for assessing future operational carbon trajectories and mitigation potential in commercial buildings across different economies.** Unlike studies focused on historical emission changes or individual countries, this framework compares the contributions of energy intensity, economic efficiency, and the comprehensive emission factor to future mitigation potential across countries. By combining static scenario modeling with Monte Carlo-based dynamic simulation, this study further evaluates the uncertainty ranges of carbon peaks, peak time, and long-term carbon lock-in levels rather than relying only on deterministic trajectories. The results help identify carbon peak and lock-in patterns across countries, clarify the main sources and end-use priorities of future mitigation potential, and inform differentiated strategies for long-term commercial building decarbonization.

The remainder of this paper is structured as follows: Section 2 reviews related studies on building emission projection and mitigation assessment. Section 3 describes the modeling framework, uncertainty simulation, and data sources. Sections 4 and 5 present the projection results, mitigation decomposition, and policy implications. Section 6 concludes the study.

## 2. Literature review

Predicting future operational carbon emissions and assessing emission reduction potential are important foundations for identifying decarbonization pathways in the building sector. Existing studies have used different models to simulate future building emissions and evaluate the effects of mitigation strategies. Based on their modeling approaches, these methods can generally be divided into top-down and bottom-up approaches [11]. Bottom-up approaches estimate emissions from detailed building-level information [12], such as building stocks, technical characteristics, and energy demand [13, 14]. Top-down approaches link carbon emissions with macrolevel drivers, such as population, economic development, energy intensity, and emission factors, and are often used to project national or sectoral emission trends [15, 16].

Representative top-down approaches include the environmental Kuznets curve (EKC) [17] and models derived from the impact of population, affluence, and technology (IPAT) framework [18]. Ghazouani [19] applied the EKC model to examine the relationship between globalization and carbon emissions in Brazil, Russia, India, China, and South Africa while also projecting future emission trends. However, EKC-based models are often limited in terms of identifying specific peak times and peak values. Therefore, IPAT-related models have been increasingly used in emission projection studies. Rokhmawati, et al. [20] combined the Stochastic Impacts by Regression on Population, Affluence, and Technology (STIRPAT) model with the EKC model to predict future carbon emission trends in Indonesia and discuss the role of carbon tax mechanisms. Tan, et al. [21] used an IPAT-based framework to quantify residential building energy consumption and carbon emissions in China, identify key driving factors, and assess future emission peaks through static and dynamic scenario simulations. On the basis of IPAT logic, the Kaya identity further decomposes carbon emissions into population, economic development, energy intensity, and emission factors [22-24]. With a concise structure and clear explanatory logic, the Kaya identity has been widely used in carbon emission driver analysis and scenario projection studies [25-27]. Beyond deterministic projections, dynamic simulations have been introduced to better reflect uncertainty in future emissions [28-30]. Monte Carlo simulation is a commonly used approach for incorporating parameter uncertainty into scenario analysis, as it can generate a range of possible future outcomes rather than a single deterministic trajectory [31]. These studies indicate that top-down models and uncertainty-based simulations provide useful tools for linking macrolevel drivers with future emission trajectories.

In addition to emission projection, assessing emission reduction potential is essential for identifying feasible pathways toward building decarbonization [32, 33]. Scenario analysis has been widely used to quantify future mitigation potential by comparing emission outcomes under different development assumptions [34, 35]. Sun, et al. [36] combined the Kaya identity and the logarithmic mean Divisia index method to evaluate historical emission changes in China's building sector and further used scenario analysis to predict future emission trends and reduction potential, revealing that mitigation potential differs across building types and is particularly high for commercial buildings. Li, et al. [37] combined the long-range energy alternative planning model with system dynamics to predict building energy consumption and carbon emissions in China from 2021 to 2060, emphasizing the role of green technologies in achieving carbon peaking. Andrews, et al. [38] developed a public building mitigation model and evaluated the carbon reduction potential of public buildings in 15 cities in the United States from 2024 to 2050, indicating that improving building performance standards and related policies is necessary to achieve an 80% reduction target by 2050. These studies show that scenario analysis is useful for estimating future mitigation potential under different assumptions. To translate mitigation potential into feasible decarbonization pathways, it is necessary to further identify the contributions of key driving factors and end uses. However, existing evidence is still derived mainly from specific countries, regions, or cities, and the understanding of future mitigation sources and policy priorities in commercial building operations from a broader cross-country perspective remains limited.

The literature on building carbon emissions has provided important support for emission projections, emission reduction potential assessment, and policy pathway design. **However**, most existing studies still focus on specific countries, regions, cities, or overall building sectors, while comprehensive cross-country assessments of future operational carbon trajectories and emission reduction potential in commercial buildings remain limited. In addition, future emission projection, uncertainty assessment, mitigation potential quantification, and driving-factor decomposition have rarely been integrated within a consistent cross-country framework, which limits the systematic identification of mitigation patterns and the translation of reduction sources into policy-oriented strategies. Therefore, further research is needed to evaluate future commercial building emissions and mitigation pathways under a consistent cross-country framework.

**To address these gaps,** this study focuses on commercial building operations in 16 countries and develops a Kaya-based model to project future operational carbon trajectories from 2024 to 2060. Monte Carlo simulation is further introduced to assess uncertainties in carbon peaks, peak times, and carbon lock-in levels. On the basis of scenario simulations, this study quantifies future emission reduction potential and decomposes the contributions of key driving factors and end uses, thereby providing an analytical basis for differentiated policy strategies. The main contributions of this study are as follows:

- **This study provides a cross-country perspective on future operational carbon trajectories in commercial buildings by combining cross-country comparisons with static and dynamic simulations.** Unlike studies focusing on individual countries, regions, or the overall building sector, this study identifies differences in future mitigation pressure across 16 countries under a consistent analytical framework.
- **This study links future emission projections, emission reduction potential assessment, and driving-factor decomposition to support policy-oriented decarbonization analysis.** By quantifying mitigation potential under the decarbonization scenario and identifying the contributions of different factors and end uses, this study provides a basis for developing differentiated policy strategies for long-term commercial building decarbonization.

## 3. Materials and methods

Section 3 presents the methodological framework for projecting future operational carbon emissions of commercial buildings across countries. Section 3.1 establishes a static emission scenario model based on the Kaya identity and defines the business-as-usual and decarbonization scenarios. Section 3.2 introduces a Monte Carlo-based dynamic uncertainty simulation to characterize future carbon peaks, peak times, and carbon lock-in levels. The data sources and parameter settings are described in Section 3.3.

### 3.1. Static emission scenario modeling for commercial buildings

Operational carbon emissions from commercial buildings are affected by population change, service-sector development, commercial floor area, energy use efficiency, and energy supply structure [39]. To simulate future carbon emission trajectories of commercial buildings across countries, this study established an emission model based on the Kaya identity. Population, per capita service value added, economic efficiency, energy intensity, and the comprehensive emission factor were selected as the key parameters [5]. The model can be expressed as follows:

$$C = P \cdot \frac{G_s}{P} \cdot \frac{F}{G_s} \cdot \sum_{i=1}^{4} \frac{E_i}{F} \cdot \frac{C}{E} \tag{1}$$

where $C$ represents total operational carbon emissions from commercial buildings, $P$ is the population of each country, $G_s$ is service value added, and $\frac{G_s}{P}$ represents per capita service value added. $F$ represents the commercial building floor area, and $\frac{F}{G_s}$ represents economic efficiency, expressed as the commercial floor area required per unit of service value added. $E_i$ is the energy consumption of end use *i*, and $\frac{E_i}{F}$ represents the energy intensity of end use *i*. $E$ is the total energy consumption of commercial buildings ($E = \sum_{i=1}^{4} E_i$), and $\frac{C}{E}$ represents the comprehensive emission factor.

For simplicity, Eq. (1) can be rewritten as follows:

$$C = P \cdot r \cdot i \cdot \sum_{i=1}^{4} e_i \cdot K \quad (2)$$

where $r$ represents per capita service value added, $i$ represents economic efficiency, $e_i$ represents the energy intensity of end use $i$, and $K$ represents the comprehensive emission factor. In this study, the four end uses include space heating, space cooling, lighting, and appliances and other end uses.

Using the operational carbon emission model as the accounting framework, this study designed two scenarios to simulate the future emission trajectories of commercial buildings across countries: the business-as-usual scenario and the decarbonization scenario. The business-as-usual scenario assumes that each country continues its existing policy framework without additional strong mitigation measures. In this scenario, the commercial floor area, energy intensity, and energy structure follow historical development trends and existing energy efficiency standards, reflecting a conventional development pathway under current socioeconomic and technological conditions.

The decarbonization scenario was designed with reference to the business-as-usual scenario, assuming more substantial socioeconomic and technological improvements to further explore the mitigation potential of commercial buildings. To maintain a consistent time horizon, this scenario was set to represent a pathway toward carbon neutrality in commercial building operations by 2060. The population and per capita service value added were kept the same as those in the business-as-usual scenario because they are driven mainly by broader socioeconomic development. In contrast, economic efficiency, energy intensity, and the comprehensive emission factor were set at lower levels than those in the business-as-usual scenario, reflecting stronger policy intervention, higher energy efficiency, and a cleaner energy supply. The detailed static parameter settings for the two scenarios are provided in Appendix B.

### 3.2. Dynamic uncertainty simulation of commercial building emissions

Static scenario analysis is used not as a precise forecast of future operational carbon emissions but as a way to describe possible emission trajectories, carbon peak or carbon lock-in patterns, and emission reduction potential under different assumptions. However, static scenario analysis relies on deterministic parameter settings and cannot fully capture future uncertainties in socioeconomic development, technological progress, energy supply, and external shocks [40]. Therefore, a Monte Carlo simulation was further introduced to improve the flexibility of the emission projection [31].

To introduce uncertainty into the static scenario model, random disturbance terms were assigned to the key parameters, including population, per capita service value added, economic efficiency, energy intensity, and the comprehensive emission factor. The dynamic value of each parameter can be expressed as follows:

$$X^{dynamic} = X^{static} \cdot \left(1 + \omega_X \cdot \frac{T - 2023}{2060 - 2023}\right), \quad \omega_X \sim N(0, \sigma_X^2) \tag{3}$$

where $X$ represents $P$, $r$, $i$, $e_i$, or $K$. $X^{static}$ is the static value of parameter $X$, and $X^{dynamic}$ is the corresponding dynamic value after considering parameter uncertainty. $T$ represents the target year during 2024–2060. Therefore, Eq. (3) describes how each parameter is adjusted in the dynamic emission model after uncertainty is incorporated.

According to the central limit theorem, when the number of samples is sufficiently large, the sample mean of a random variable tends to follow a normal distribution. Therefore, the random parameter $\omega$ for each variable was assumed to follow a normal distribution $N(0, \sigma^2)$. The standard deviation (SD) σ was determined based on the approximately 95.45% probability coverage of the ±2σ interval under the assumed normal distribution. Specifically, the value of $\sigma$ was set so that the maximum and minimum values of each variable under the business-as-usual and decarbonization scenarios were covered by the interval $\pm 2\sigma$, corresponding to a probability range of 95.45% under the normal distribution. The detailed uncertainty settings for each country are provided in Appendix C.

After the distributions of random parameters were determined, Monte Carlo simulation was used to conduct repeated sampling. The sampled parameters were introduced into the operational carbon emission model, and the static emission results were transformed into dynamic emission results:

$$\frac{C}{(static)} \xrightarrow[]{\substack{Random \\ parameters}} \frac{C \cdot (1 + \omega)}{(dynamic)}, \quad \omega \sim N(0, \sigma_C^2) \tag{4}$$

where $C^{static}$ represents the emission result calculated from the static scenario parameters and $C^{dynamic}$ represents the corresponding emission result after considering parameter uncertainty. In this study, 100,000 Monte Carlo simulations were performed for each country to obtain the probability distributions of future operational carbon emissions. The simulation outputs were subsequently used to characterize the distributions of emission peak magnitudes and peak years, as well as the future carbon lock-in levels of commercial building operations across countries.

### 3.3. Datasets

To ensure data completeness and cross-country comparability in simulating future operational carbon emissions, this study compiled relevant socioeconomic, building, energy, and emission data for 16 countries. Population projections were obtained mainly from the United Nations World Population Prospects 2024. Service value added was calculated on the basis of the service-sector share of gross domestic product (GDP) and GDP data from the World Bank (data.worldbank.org) and national statistical sources. Commercial building floor area was collected from the International Energy Agency and supplemented by national statistical yearbooks, official building sector reports, and previous studies. Energy intensity and emission factor parameters were determined with reference to the International Energy Agency and the Global Building Emissions Database (GLOBE, https://globe2060.org/). Detailed data sources and parameter settings are provided in the appendices.

## 4. Results

### 4.1. Peaking dynamics and carbon lock-in levels of future operational emissions

The historical and projected carbon emission trajectories of commercial building operations in the 16 countries under the business-as-usual scenario are presented in Fig. 1. The gray lines depict historical emissions from 2000 to 2023, whereas the brown lines indicate the static trajectories simulated for 2024–2060. Considering the uncertainties in the impact parameters, the potential fluctuation ranges of carbon emissions were further quantified. The green bands represent these ranges under different uncertainty levels, corresponding to ±1, ±2, and ±3 SD.

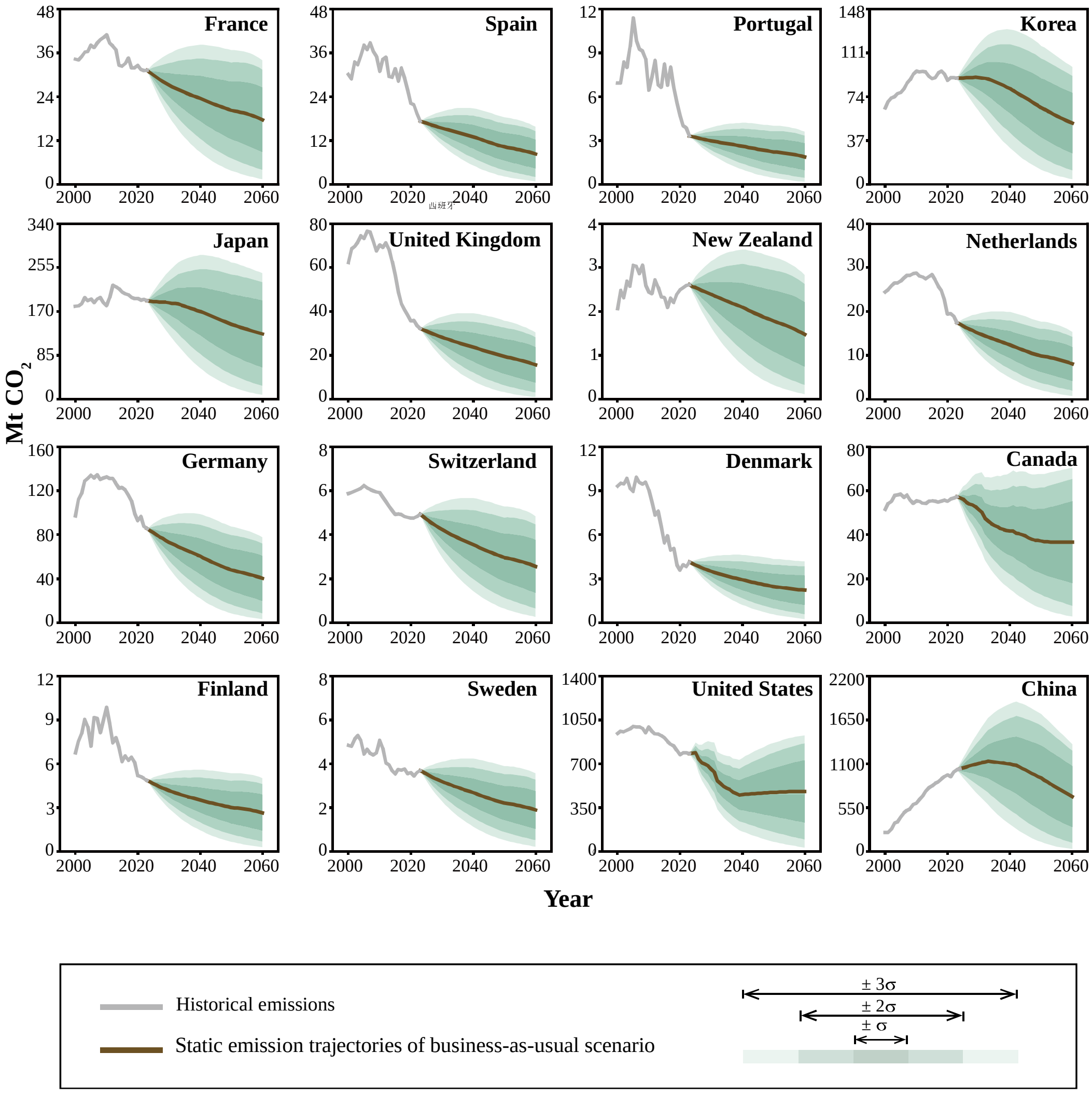

**Fig. 1.** Static operational carbon emission trajectories and dynamic emission ranges for commercial buildings in various countries under the business-as-usual scenario up to 2060.

At the national level, future emissions in China show an inverted U-shaped trajectory under the business-as-usual scenario and are projected to reach an emission peak of 1128 million tonnes of carbon dioxide ($MtCO_2$) in 2033. The trajectory for South Korea also shows an inverted U-shaped pattern. However, unlike China, South Korea's commercial building emissions peaked earlier in 2018, with a peak value of 96 $MtCO_2$. In the United States and Canada, emissions will gradually enter a carbon lock-in state. Specifically, emissions in the United States are projected to remain locked at approximately 460 $MtCO_2$, whereas those in Canada will stabilize at approximately 38 $MtCO_2$. For the other countries, namely, France, Spain, Portugal, the United Kingdom, New Zealand, the Netherlands, Germany, Switzerland, Denmark, Finland, Sweden, and Japan, emissions generally decline over time, but their rates of decline gradually slow, suggesting a possible transition toward carbon lock-in in the later projection period.

The distributions of the potential emission peak and peak time in China and the carbon lock-in levels in the other countries based on 100,000 Monte Carlo simulations under the business-as-usual scenario are shown in Fig. 2. These results complement the static trajectories in Fig. 1 by capturing the uncertainties in future emission changes.

As shown in Fig. 2, the red and green distributions indicate the potential emission peak and corresponding peak time of commercial building operations in China, respectively. Based on the Monte Carlo simulations, China's commercial building operational carbon emissions are projected to peak in 2033 (±5 years) at 1150 (±120) $MtCO_2$, with the reported uncertainties representing one standard deviation.

For the other countries, the results mainly reflect the projected carbon lock-in levels of commercial building operations. The high lock-in group includes the United States, Japan, South Korea, Germany, and Canada. Among them, the United States and Japan show the most prominent lock-in levels, reaching 470 (±89) $MtCO_2$ and 141 (±31) $MtCO_2$, respectively, while the remaining countries in this group are projected to range from 39 to 63 $MtCO_2$. These countries show considerably higher lock-in levels than the others do, indicating greater long-term mitigation pressure. The medium lock-in group includes France, the United Kingdom, Spain, and the Netherlands, whose lock-in levels range from 11 to 21 $MtCO_2$. The low lock-in group includes mainly Finland, Switzerland, Denmark,

Portugal, Sweden, and New Zealand, whose lock-in levels remain approximately 3 $MtCO_2$ or lower. Overall, the carbon lock-in levels are concentrated in a small number of countries, while most countries show relatively low or medium lock-in levels.

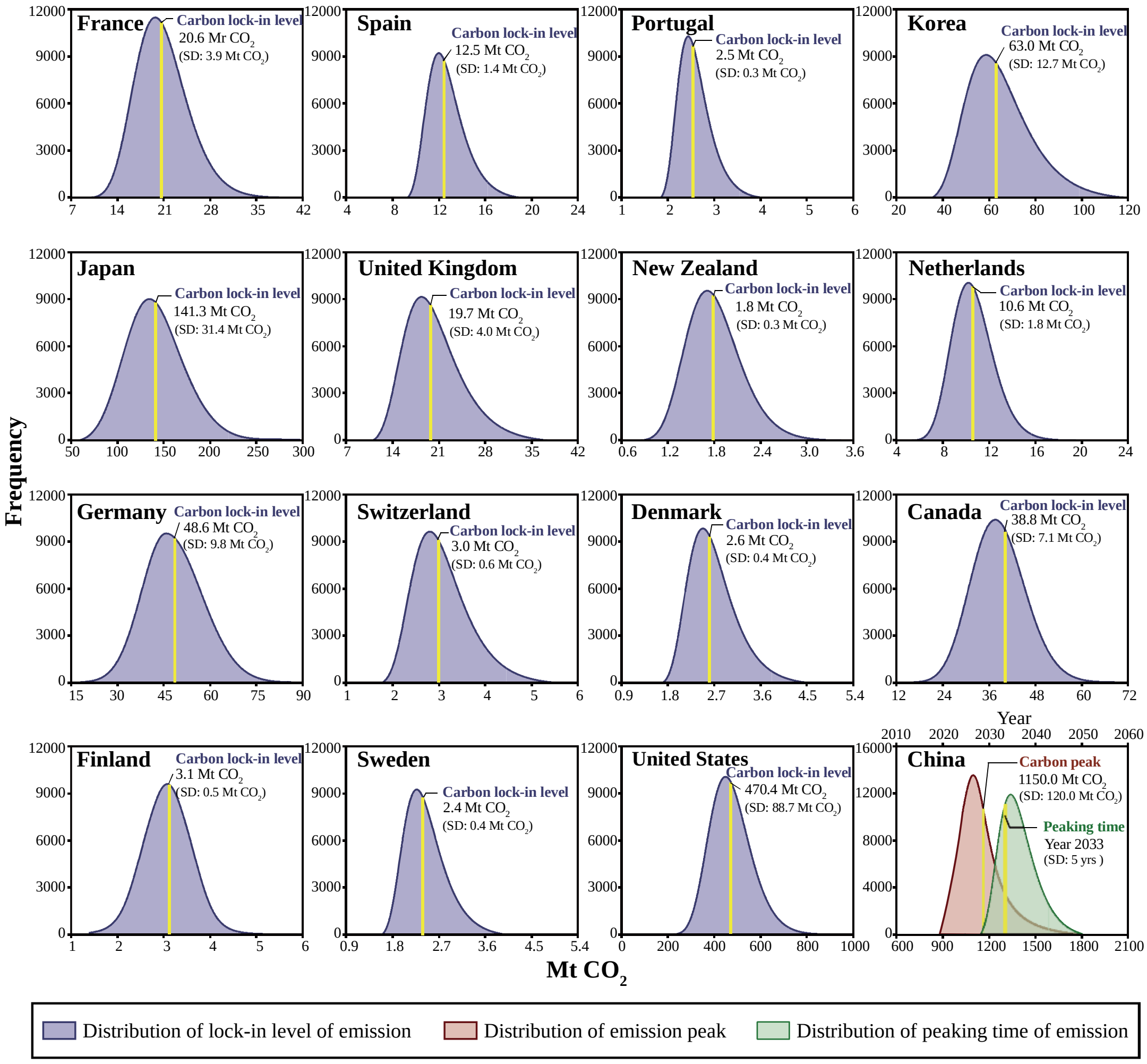

**Fig. 2.** Operational carbon peak or lock-in distributions in commercial buildings in various countries. **Note:** The purple distributions represent the lock-in levels of carbon emissions in various countries, the distributions in red represent the emission peaks in China, and the distributions in green represent the emission peak time in China.

Carbon lock-in levels differ markedly across regions. North American countries, especially the United States and Canada, are projected to remain at relatively high lock-in levels, with the United States ranking first among all countries. Japan and South Korea in East Asia also show relatively prominent lock-in levels. In contrast, European countries such as France, the United Kingdom, Spain, and the Netherlands fall mainly within the medium lock-in group, showing relatively comparable lock-in levels. Nordic countries, including Sweden, Denmark, and Finland, together with New Zealand, show the lowest projected lock-in levels.

Overall, the static trajectory analysis in Fig. 1 and the Monte Carlo-based assessment of carbon peaks, peak times, and carbon lock-in levels in Fig. 2 clarify the future operational carbon trajectories of commercial buildings across countries, thereby addressing Question 1 in Section 1.

### 4.2. Spatiotemporal evolution of future emission reduction potential

The 2024 baseline emissions and the spatiotemporal distribution of the emission reduction potential in commercial building operations across the 16 countries from 2030 to 2060 are presented in Fig. 3. The 2024 baseline map shows that operational carbon emissions are highly concentrated in a small number of countries. China and the United States have the highest baseline emissions, reaching 1035 $MtCO_2$ and 789 $MtCO_2$, respectively, followed by Japan, South Korea, Germany, and Canada. Together, China and the United States account for 77.1% of the total baseline emissions in the 16 countries, indicating a highly uneven baseline emission pattern.

At the temporal level, the total emission reduction potential is projected to increase substantially from 352 $MtCO_2$ in 2030 to 799 $MtCO_2$ in 2040 and 1119 $MtCO_2$ in 2050. After 2050, the growth rate slows markedly, and the total reduction potential stabilizes at approximately 1123 $MtCO_2$ by 2060. This finding indicates that the emission reduction potential of commercial building operations expands rapidly before 2050 but gradually approaches a plateau thereafter.

At the national level, China consistently has the greatest emission reduction potential for 2030–2060. Its reduction potential increases from 196 $MtCO_2$ in 2030 to 467 $MtCO_2$ in 2040 and 613 $MtCO_2$ in 2050 before decreasing slightly to 539 $MtCO_2$ in 2060. The United States ranks second in all periods, with its reduction potential increasing continuously from 80 $MtCO_2$ in 2030 to 390 $MtCO_2$ in 2060. Japan also maintains a relatively high reduction potential, increasing from 22 $MtCO_2$ in 2030 to 65 $MtCO_2$ in 2060. In addition, South Korea,

Germany, and Canada form the second tier of countries with relatively high reduction potential. By 2060, their reduction potential will reach 33 $MtCO_2$, 32 $MtCO_2$, and 24 $MtCO_2$, respectively.

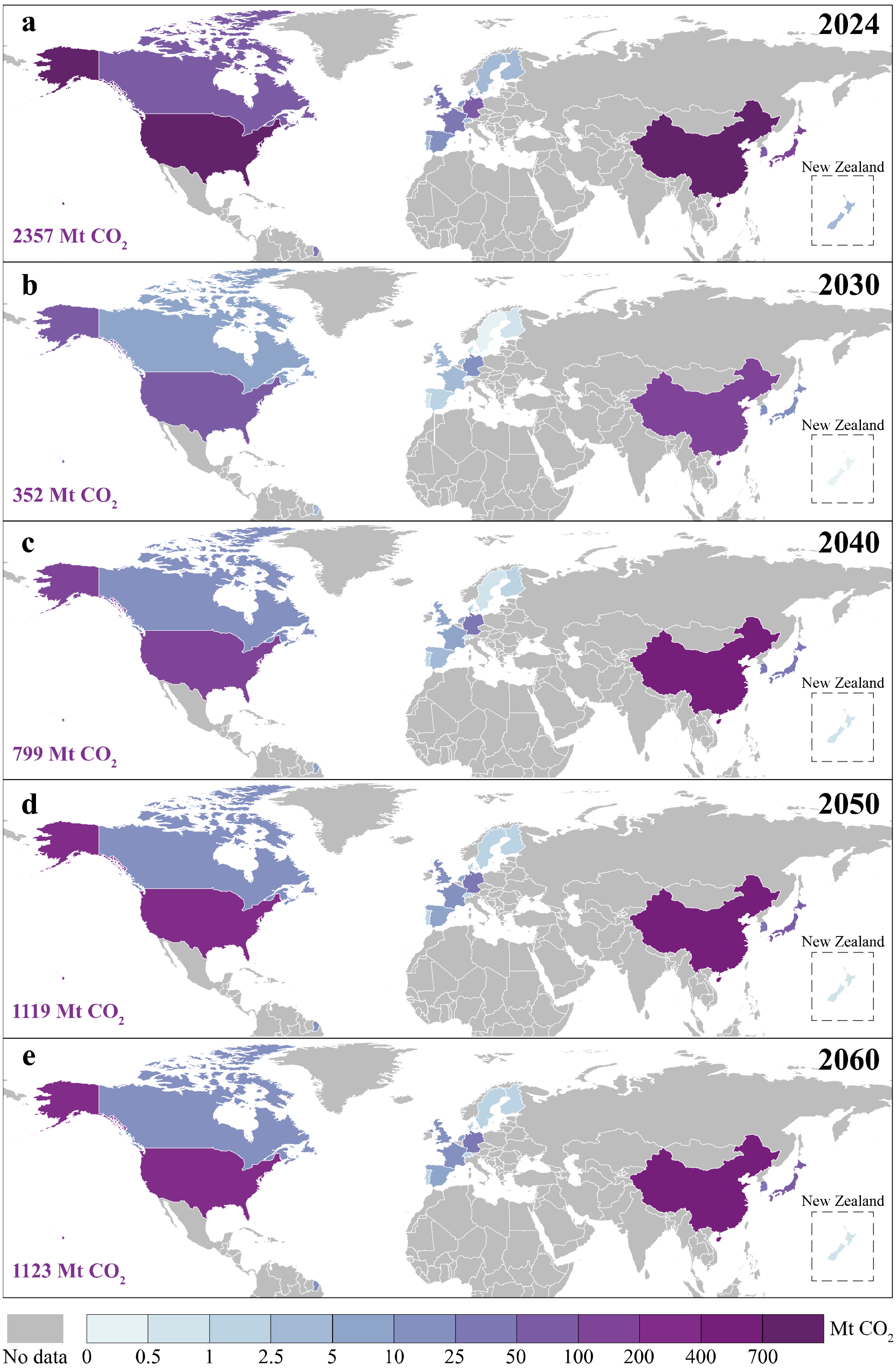

**Fig. 3.** Spatiotemporal distribution of the emission reduction potential of commercial buildings across various countries (2024–2060). **Note:** The maps illustrate the baseline emissions in 2024 and the emission reduction potential in 2030, 2040, 2050, and 2060.

The spatial distribution further indicates that the emission reduction potential is highly concentrated among a few major emitters. The top three countries, namely, China, the United States, and Japan, contribute 84.7%–88.9% of the total reduction potential during 2030–2060. When South Korea and Germany are included, the share of the top five countries remains above 93.0% in all periods. In contrast, countries outside the top five show much lower absolute reduction potential, especially New Zealand, Sweden, Switzerland, Portugal, Denmark, and Finland, where the emission reduction potential remains below 2 $MtCO_2$ in 2060. Overall, Fig. 3 indicates that the future emission reduction potential of commercial building operations is unevenly distributed across countries, with major emitters contributing most of the total reduction potential, while the total reduction potential tends to stabilize after 2050.

# 5. Discussion

After the future operational carbon trajectories, carbon peak or lock-in patterns, and the emission reduction potential of commercial buildings are identified, it is necessary to further discuss the sources of mitigation potential to clarify how future decarbonization can be achieved across countries. Therefore, Section 5.1 decomposes the emission reduction potential into contributions from three key driving factors: energy intensity, economic efficiency, and the comprehensive emission factor. In Section 5.2, the emission reduction potential from energy intensity reduction is further decomposed by end use to identify differentiated building-side mitigation priorities. Finally, Section 5.3 presents policy strategies for promoting commercial building decarbonization according to the dominant end-use characteristics and the absolute scale of mitigation potential in each country.

## 5.1. Decomposition of driving factors and their contributions to emission reduction potential

Fig. 4 further decomposes the emission reduction potential of commercial building operations in 2060 into contributions from three driving factors: energy intensity, economic efficiency, and the comprehensive emission factor. Building on the emission reduction potential identified in Fig. 3, Fig. 4 focuses on the sources of the gap between the business-as-usual and decarbonization scenarios in 2060 and explains why the contributions of the driving factors differ across countries.

At the factor level, the decline in the comprehensive emission factor is the dominant contributor in most countries. Its contribution reaches 72.6% in the United States, 62.5% in Portugal, 62.3% in Germany, 61.7% in Denmark, 60.8% in Finland, and 60.0% in China. Similar patterns are also observed in South Korea, the Netherlands, France, Spain, and Canada, where the comprehensive emission factor contributes more than half of the total emission reduction potential. This pattern suggests that, in most countries, a large part of the reduction potential is associated with cleaner energy use rather than only with reductions in building-side energy demand. In this sense, the comprehensive emission factor links commercial building decarbonization with the carbon intensity of electricity, heat, and fuel supply, highlighting the importance of energy-system decarbonization.

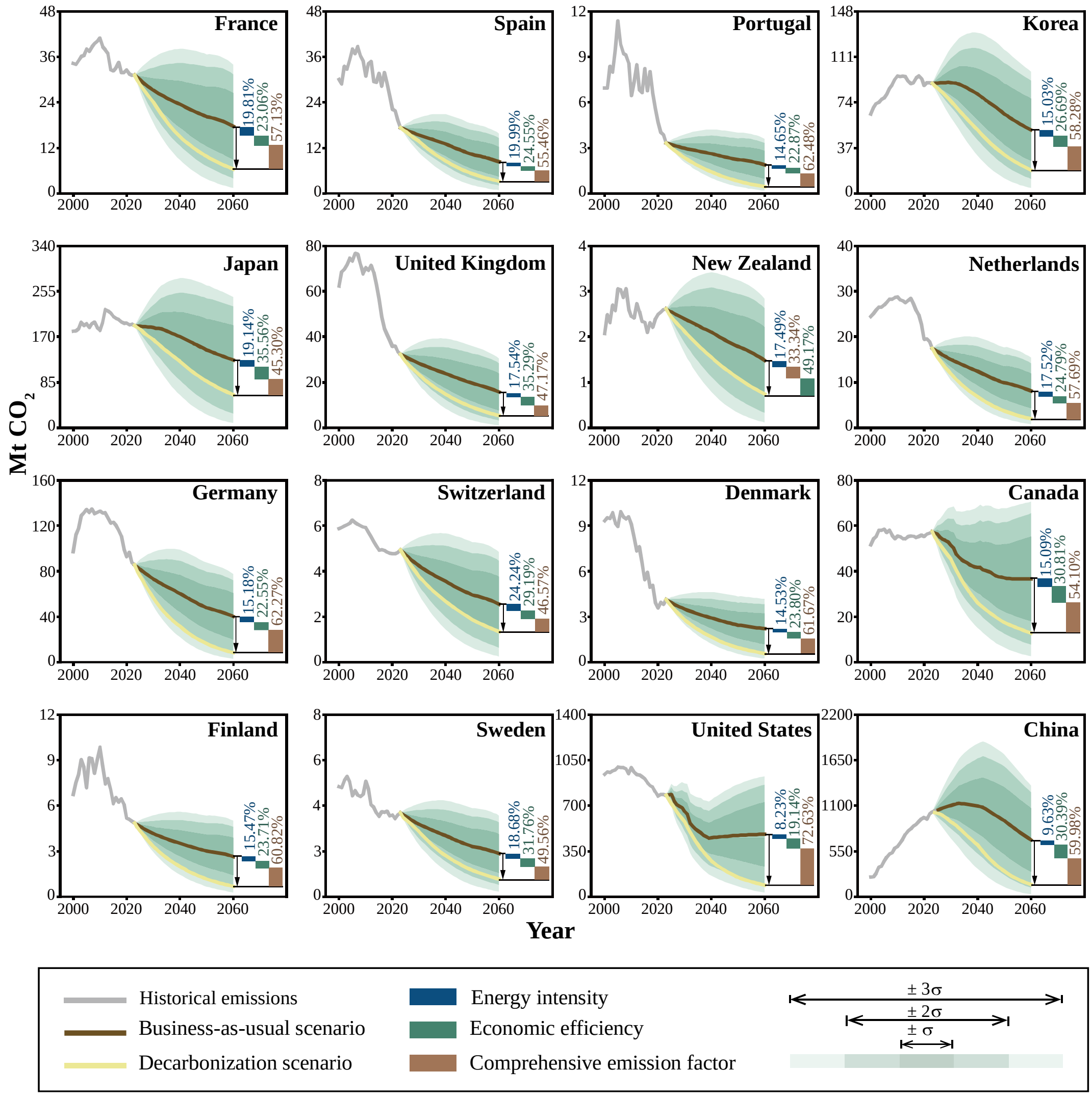

**Fig. 4.** Emission reduction potential and decomposed driving factors of commercial buildings across various countries by 2060.

Economic efficiency generally provides the second-largest contribution in most countries. Its contribution is relatively high in Japan, the United Kingdom, Sweden, Canada, Switzerland, and China, ranging from approximately 30.0% to 36.0%. The contribution of economic efficiency indicates that part of the emission reduction potential is related to whether service-sector output can increase without a proportional expansion of commercial floor area. In other words, improving economic efficiency means supporting more service activities with less additional building space, thereby reducing operational emissions associated with excess floor area, idle space, or low-intensity space use. New Zealand is the main exception to the general pattern. Unlike most other countries, its emission reduction potential is driven mainly by economic efficiency, which contributes 49.2%, with the comprehensive emission factor ranking second. This reflects the cross-country heterogeneity in the drivers of commercial building decarbonization.

Energy intensity also contributes to the emission reduction potential of commercial building operations. Its contribution ranges from 8.2% in the United States to 24.2% in Switzerland, with relatively high shares also observed in Spain, France, Japan, and Sweden. Energy intensity is closely related to technological improvement in commercial building systems and has considerable decarbonization potential, making it a key factor for understanding building-side mitigation opportunities. Overall, as shown in Fig. 4, by 2060, the emission reduction potential of commercial buildings is driven mainly by the decline in the comprehensive emission factor in most countries, while economic efficiency provides an important additional contribution, and energy intensity serves as a common building-side mitigation pathway.

### 5.2. End-use decomposition of energy intensity and contributions to emission reduction potential

To further identify end-use contributions to energy-intensity-related mitigation potential, this section decomposes the projected emission reduction potential from energy intensity reduction into four end uses: space heating, space cooling, lighting, and appliances and other end uses. As shown in Fig. 5, the map presents the country-level emission reduction potential attributable to energy intensity reduction in 2060, while the waterfall charts further show the contribution shares of the four end uses.

At the national level, the emission reduction potential from energy intensity reduction is highly uneven across countries. China has the greatest potential, reaching 52 $MtCO_2$ in 2060, followed by the United States at 32 $MtCO_2$ and Japan at 12 $MtCO_2$. South Korea, Germany, Canada, France, the United Kingdom, the Netherlands, and Spain form the second tier, with potentials ranging from 1 to 5 $MtCO_2$. In contrast, Denmark, Finland, Switzerland, Portugal, Sweden, and New Zealand show much lower absolute potentials, all of which are less than 0.3 $MtCO_2$. This pattern is broadly consistent with the overall emission reduction potential shown in Fig. 3, indicating that countries with larger commercial building emission scales also tend to have greater absolute mitigation potential from energy intensity improvement.

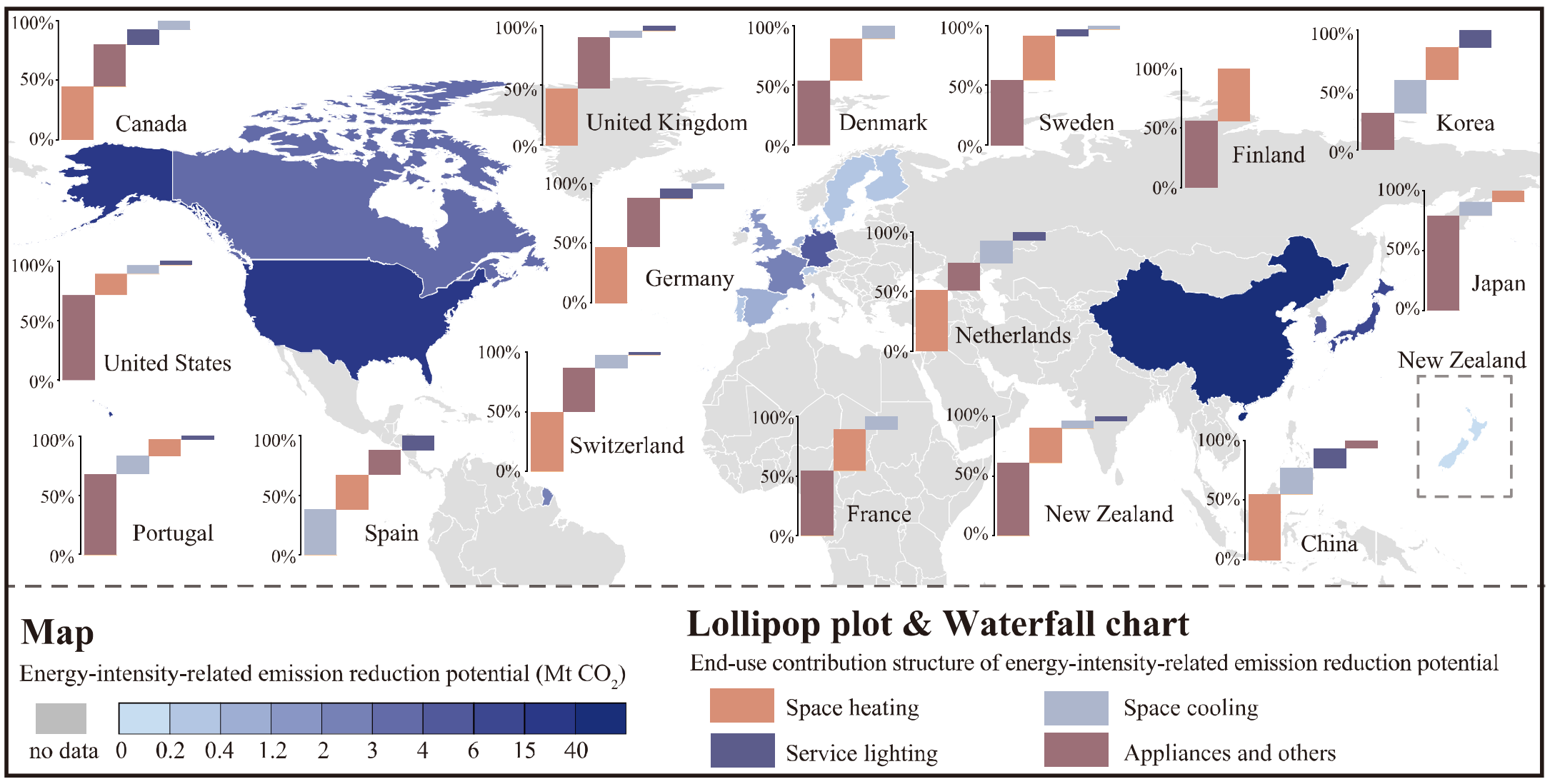

**Fig. 5.** Emission reduction potential attributable to energy intensity reduction and its end-use contribution shares in commercial buildings across various countries by 2060. **Note:** End-use categories vary across countries because of data availability. Service-lighting information was unavailable for France, Denmark, and Japan, and Finland was represented using only space heating and appliances and other end uses.

End-use contributions differ substantially across countries, providing a basis for identifying country-specific mitigation priorities. In China, space heating is the greatest contributor to the emission reduction potential from energy intensity reduction, accounting for 55.0%, followed by space cooling (22.0%), lighting (16.8%), and appliances and other end uses (6.1%). These findings suggest that improvements in heating efficiency remain the most important mitigation pathway for commercial buildings in China, while cooling and lighting also provide considerable additional potential. Similar heating-first patterns are observed in Germany, the United Kingdom, Canada, the Netherlands, and Switzerland, where space heating contributes 46.8%, 47.4%, 44.3%, 51.4%, and 49.9%, respectively. These countries therefore exhibit relatively high mitigation potential associated with heating energy intensity reduction.

In contrast, appliances and other end uses dominate the contribution structure in several countries. The share of appliances and other end uses reaches 71.2% in the United States and 79.1% in Japan. Similar patterns are also observed in Portugal, New Zealand, Finland, Sweden, France, and Denmark, where appliances and other end uses contribute 67.9%, 60.9%, 56.0%, 54.4%, 54.2%, and 54.2%, respectively. These results indicate that appliances and other end uses can serve as the main source of mitigation potential from energy intensity improvement in many countries. Space cooling also makes a notable contribution to mitigation potential in some countries, especially Spain (38.0%), South Korea (28.0%), China (22.0%), and the Netherlands (19.1%). These findings indicate that cooling efficiency may become an important mitigation pathway in countries with stronger cooling demand or greater growth potential in terms of cooling energy use. South Korea shows a relatively balanced structure among appliances and other end uses, space cooling, and space heating, suggesting that its mitigation potential is more closely related to combined improvements in heating, ventilation, and air conditioning (HVAC) systems and equipment efficiency. Lighting contributes less than heating, cooling, and appliances in most countries, but it still accounts for relatively high shares in China (16.8%), South Korea (14.1%), Canada (12.4%), and Spain (12.0%).

The absolute emission reduction potential of each end use across countries is further ranked in Fig. 6, thereby distinguishing the relative contribution structure from the absolute mitigation scale. With respect to space heating, China ranks first, with a reduction potential of 29 Mt$CO_2$, far exceeding that of the United States (6 Mt$CO_2$), Germany (2 Mt$CO_2$), Canada (2 Mt$CO_2$), South Korea (1 Mt$CO_2$), and Japan (1 Mt$CO_2$). With respect to space cooling, China also has the greatest potential, reaching 11 Mt$CO_2$, followed by the United States (2 Mt$CO_2$), Japan (2 Mt$CO_2$), and South Korea (1 Mt$CO_2$). This finding indicates that although the contribution share of cooling is not always dominant, its absolute mitigation potential can be considerable in countries with large commercial building stocks.

With respect to lighting, China again ranks first, with a reduction potential of 9 Mt$CO_2$, followed by the United States (1 Mt$CO_2$), South Korea (1 Mt$CO_2$), Canada (0.4 Mt$CO_2$), and Germany (0.4 Mt$CO_2$). Lighting contributions could not be evaluated for Denmark, France, and Japan because of data limitations, while Finland's decomposition included only space heating and appliances and other end uses. With respect to appliances and other end uses, the United States has the greatest reduction potential, reaching 23 $MtCO_2$, followed by Japan (10 $MtCO_2$) and China (3 $MtCO_2$). These results differ from those of the other three end uses, for which China consistently ranks first, and suggest that the United States and Japan have particularly large mitigation potential from improving the efficiency of appliances and other end uses in commercial buildings.

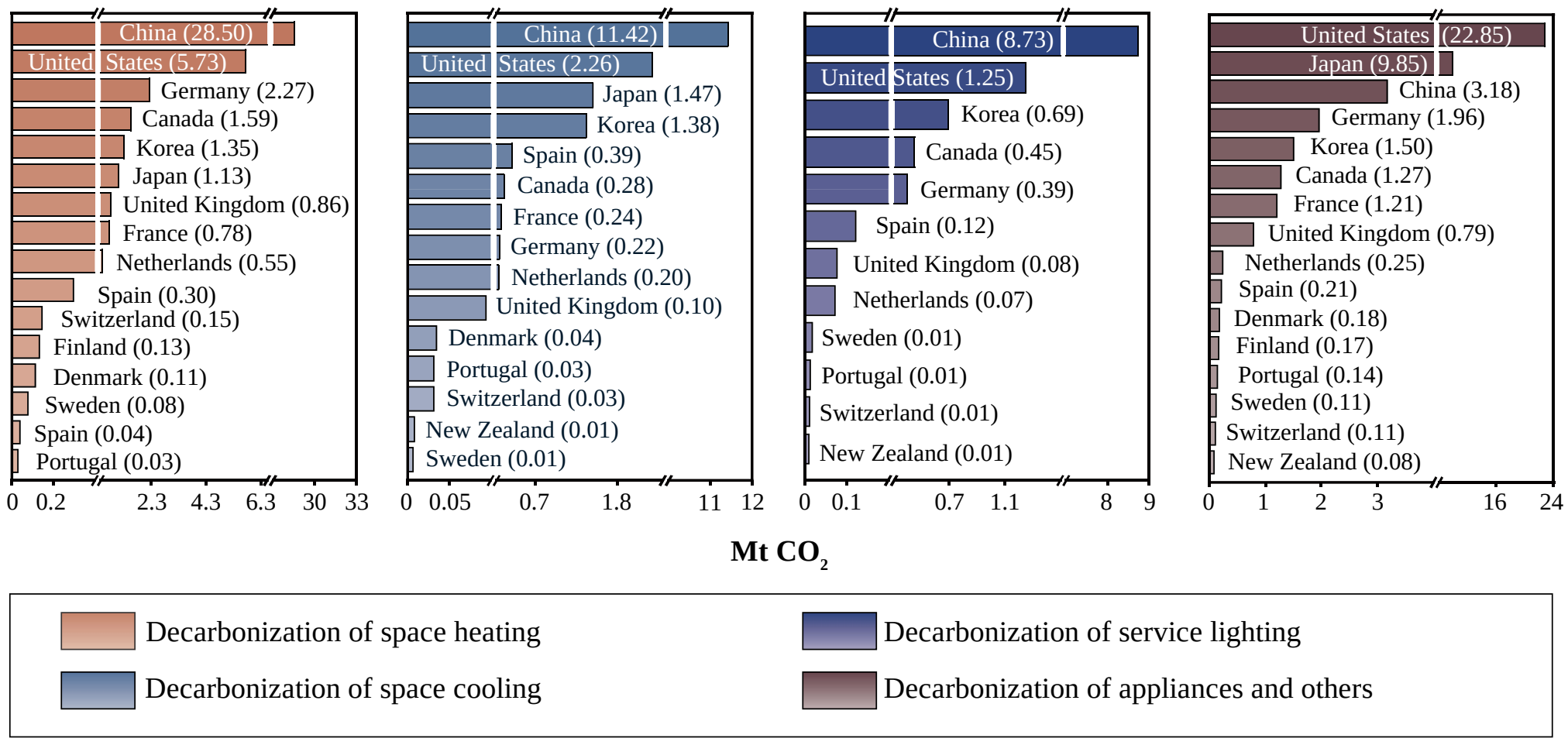

**Fig. 6.** Country ranking of end-use-specific emission reduction potential in commercial buildings by 2060. **Note:** End-use categories vary across countries because of data availability. Service-lighting information was unavailable for France, Denmark, and Japan, and Finland was represented using only space heating and appliances and other end uses.

At the end-use level, energy intensity reduction provides a common but heterogeneous mitigation pathway for commercial buildings across countries. At the aggregate level, the two largest contributors are space heating and the category comprising appliances and other end uses, while space cooling and lighting provide additional reduction potential in specific countries. China has the greatest absolute reduction potential in space heating, space cooling, and lighting, whereas the United States has the greatest potential in appliances and other end uses. These differences in end-use contributions suggest that future energy efficiency strategies for commercial buildings should not rely on a single pathway. Instead, priorities for each country should be developed according to the dominant end uses and the absolute scale of the associated mitigation potential.

Overall, the spatiotemporal distribution of the emission reduction potential in Fig. 3,

the driving-factor decomposition in Fig. 4, and the end-use decomposition and country ranking of the emission reduction potential from energy intensity reduction in Figs. 5 and 6 identify future mitigation potential in commercial building operations, its key driving factors, and its major end-use contributors, thereby addressing Question 2 in Section 1.

### 5.3. Policy recommendations

Sections 5.1 and 5.2 show that the emission reduction potential of commercial building operations is shaped mainly by the comprehensive emission factor, economic efficiency, and energy intensity. Although all three factors are important, this section focuses on energy intensity because it is more directly related to technological improvement within commercial buildings. Therefore, this section translates the end-use decomposition of energy-intensity-related mitigation potential into differentiated policy pathways. Based on the dominant end-use characteristics identified in Section 5.2, four differentiated decarbonization pathways are proposed for the 16 countries, as shown in Fig. 7.

For China, Germany, the United Kingdom, Canada, the Netherlands, and Switzerland, the relatively high contribution of space heating indicates a heating-first pathway for commercial building decarbonization. In these countries, improving heating efficiency should be a priority. Policies should first reduce heat losses through the building envelope. Measures such as improved insulation [41], high-performance windows [42], and better airtightness can lower heating loads in commercial buildings [43]. Thermal storage and smart thermostats should also be adopted to improve the flexibility of heating loads and reduce peak demand [44, 45]. As the power supply becomes cleaner, heating systems that rely on fossil fuels should be gradually replaced with electric heating technologies, such as heat pumps [46]. Therefore, countries in this group should combine lower heating demand, load flexibility, and the transition from fossil fuel heating to electric heating to promote commercial building decarbonization.

Spain represents a cooling-first pathway, as space cooling is the greatest contributor to its potential for reducing energy intensity. These findings indicate that cooling demand control and cooling system efficiency are the main priorities for commercial building decarbonization. For countries with high or increasing cooling demand, controlling solar heat gains is essential for reducing cooling loads [47]. External shading, optimized energy-efficient glazing, and solar control devices can limit excessive heat gains in commercial buildings and reduce the need for mechanical cooling [48-50]. High-efficiency chillers should also be promoted to reduce electricity use in cooling systems [51]. In addition, passive cooling strategies should be incorporated into the design stage of new buildings and retrofit projects [52, 53], including natural ventilation [54], solar control, and climate-responsive building design [55]. Although this pathway is most directly applicable to Spain, it also has implications for countries with notable cooling contributions, such as South Korea, China, and the Netherlands.

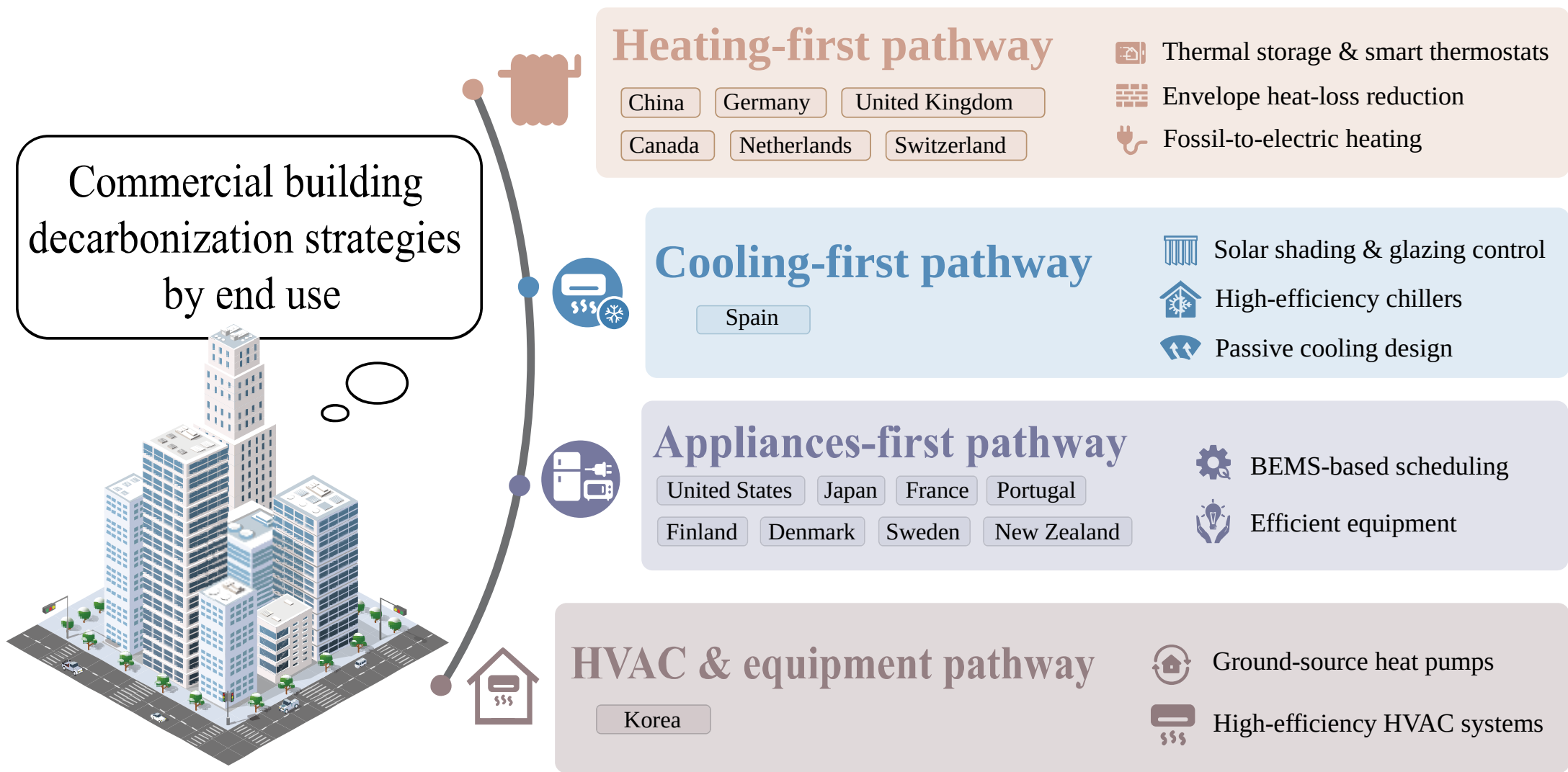

**Fig. 7.** End-use pathways for commercial building decarbonization across various countries. Note: BEMS denotes the building energy management system, and HVAC denotes heating, ventilation, and air conditioning.

In the United States, Japan, France, Portugal, Finland, Denmark, Sweden, and New Zealand, the dominance of appliances and other end uses points to an appliance-first pathway. Therefore, mitigation policies in these countries should extend beyond heating and cooling systems. High-efficiency equipment should be promoted in offices, retail spaces, and service facilities to reduce electricity demand from daily operations and other end uses [56, 57]. In addition, scheduling supported by building energy management systems (BEMS) can optimize operating hours, reduce idle energy consumption [58], and coordinate equipment loads with overall building operation [59, 60]. For countries such as the United States and Japan, where the absolute mitigation potential from appliances and other end uses is particularly large, equipment efficiency and operational management should become key policy priorities.

South Korea differs from the other countries because its mitigation potential is distributed more evenly across HVAC systems and equipment-related end uses. Therefore, mitigation policies should improve HVAC system performance and equipment efficiency together rather than focusing on one dominant end use. Ground-source heat pumps should be promoted to provide efficient heating and cooling services in commercial buildings [61]. In addition, high-efficiency HVAC systems can improve the overall performance of indoor climate control and reduce energy demand from heating and cooling operations [62]. Equipment efficiency and system scheduling should also be strengthened in an integrated manner so that efficiency gains in one end use are not offset by higher demand in another.

On the basis of the end-use decomposition results, Fig. 7 translates the dominant mitigation characteristics into differentiated decarbonization pathways. The heating-first pathway emphasizes heating demand reduction and electrified heating; the cooling-first pathway focuses on cooling demand control and efficient cooling systems; the appliance-first pathway highlights efficient equipment and operational scheduling; and the HVAC and equipment pathway emphasizes integrated system improvement. These pathways are consistent with the results in Section 5.2 and provide a basis for designing commercial building decarbonization strategies according to the dominant end uses and the absolute scale of mitigation potential in each country.

Overall, the end-use decomposition and ranking results in Figs. 5 and 6, together with the differentiated policy pathways proposed in Fig. 7, identify mitigation priorities across countries and end uses, thereby addressing Question 3 in Section 1.

## 6. Conclusion

In this study, an analytical framework integrating Kaya-based projection and Monte Carlo uncertainty analysis was developed to assess operational emission trajectories and mitigation potential from 2024 to 2060 for commercial buildings in 16 countries. First, an operational carbon emission projection model was established to compare business-as-usual and decarbonization trajectories through static scenario analysis and to assess uncertainty in carbon peak and lock-in characteristics through dynamic stochastic simulation. Afterward, the emission reduction potential was quantified and decomposed by key driving factors and end-use energy intensity improvements. Finally, policy recommendations were proposed to support commercial building decarbonization and help translate future mitigation potential into actual emission reductions. The main findings are summarized below.

### 6.1. Main findings

- **Under the business-as-usual scenario, China's commercial building operational carbon emissions are projected to peak at 1150 (±120) $MtCO_2$ in 2033 (±5 years), while the United States shows the highest carbon lock-in level of 470 (±89) $MtCO_2$.** South Korea also shows an inverted U-shaped trajectory, but its commercial building emissions peaked in 2018, reaching 96 $MtCO_2$. For the other countries, the future emission trajectories mainly reflect different carbon lock-in levels. The high lock-in group includes the United States, Japan, South Korea, Germany, and Canada, whereas France, the United Kingdom, Spain, and the Netherlands form the medium lock-in group. Finland, Switzerland, Denmark, Portugal, Sweden, and New Zealand remain in the low lock-in group. Overall, carbon lock-in pressure is concentrated among a small number of major emitters, especially in North America and East Asia, whereas most European countries show relatively low or medium lock-in levels.
- **The total emission reduction potential of commercial building operations increases from 352 $MtCO_2$ in 2030 to 1119 $MtCO_2$ in 2050 and then stabilizes at approximately 1123 $MtCO_2$ by 2060.** China consistently shows the greatest reduction potential, increasing from 196 $MtCO_2$ in 2030 to 613 $MtCO_2$ in 2050 before decreasing slightly to 539 $MtCO_2$ in 2060. The United States ranks second in all periods, with its reduction potential reaching 390 $MtCO_2$ in 2060, followed by Japan, South Korea, Germany, and Canada. The top three countries, namely, China, the United States, and Japan, contribute 84.7%–88.9% of the total reduction potential during 2030–2060. By 2060, the decline in the comprehensive emission factor is the dominant contributor in most countries, with contributions reaching 72.6% in the United States and 60.0% in China. Economic efficiency generally provides the second-largest contribution, especially in Japan, the United Kingdom, Sweden, Canada, Switzerland, and China, while energy intensity serves as an important mitigation pathway for commercial buildings.

- **By 2060, the emission reduction potential from energy intensity reduction reaches 52 $MtCO_2$ in China, 32 $MtCO_2$ in the United States, and 12 $MtCO_2$ in Japan, indicating a highly uneven distribution across countries.** In China, space heating is the greatest contributor, accounting for 55.0%, followed by space cooling, lighting, and appliances and other end uses. Similar heating-first patterns are observed in Germany, the United Kingdom, Canada, the Netherlands, and Switzerland. In contrast, appliances and other end uses dominate the contribution structure in the United States, Japan, Portugal, New Zealand, Finland, Sweden, France, and Denmark. Space cooling makes a notable contribution to mitigation potential in Spain, South Korea, China, and the Netherlands, whereas lighting contributes less than the other end uses in most countries. In terms of absolute reduction potential, China ranks first in space heating, space cooling, and lighting, whereas the United States ranks first in appliances and other end uses. Therefore, future energy intensity reduction should focus on differentiated end-use priorities rather than a single pathway across countries.

### 6.2. Future work

To further improve the assessment of future operational carbon trajectories and mitigation pathways in commercial buildings, several issues need to be addressed in future research.

First, more detailed end-use classifications are needed. In this study, commercial building end uses were divided into four categories, but more refined activity-level data could help identify specific energy use scenarios and clarify their mechanisms in shaping operational carbon emissions. Second, the country sample should be further expanded. Owing to limitations in data availability and comparability, this study focused on 16 countries and did not fully cover many regions, including South America, Africa, and Southeast Asia. Future studies could include more countries with different development levels, climate conditions, and energy endowments to improve the global representativeness of the analytical framework. In addition, future research could extend the analysis from the national scale to the city scale. Considering the large differences in climate, economic development, building stock, and energy structure within and across countries, city-level and multidimensional comparisons would help identify more region-specific decarbonization pathways and inform more targeted policies for reducing operational carbon emissions from commercial buildings.

## Acknowledgment

N.Z. acknowledges support from the Energy Demand Changes Induced by Technological and Social Innovations network, an initiative coordinated by the Research Institute of Innovative Technology for the Earth and the International Institute for Applied Systems Analysis, and funded by the Ministry of Economy, Trade, and Industry, Japan.